\documentclass{article} 
\usepackage{icad2006,amssymb,amsmath,epsfig,url} 
\usepackage[utf8x]{inputenc}

\ninept 
\title{Sonification Abstraite/Sonification Concrète:\\
An `Æsthetic Perspective Space' for Classifying Auditory Displays in the Ars Musica Domain} 
\twoauthors{Paul Vickers}{Northumbria University \\
School of Computing, Engineering, \\\& Information Sciences \\ 
Pandon Building, Camden St. \\
Newcastle upon Tyne, NE2 1XE, UK\\ 
{\tt paul.vickers@unn.ac.uk}} 
{Bennett Hogg}{International Centre for Music Studies\\
University of Newcastle upon Tyne\\
School of Arts and Cultures\\
Armstrong Building\\
Newcastle upon Tyne, NE1 7RU, UK\\
{\tt bennett.hogg@newcastle.ac.uk}}
\begin{document} 
\maketitle 
\begin{abstract} 
This paper discusses æsthetic issues of sonifications and the relationships between sonification (\textit{ars informatica}) and music \& sound art (\textit{ars musica}). It is posited that many sonifications have suffered from poor internal ecological validity which makes listening more difficult, thereby resulting in poorer data extraction and inference on the part of the listener. Lessons are drawn from the electroacoustic music and musique concrète communities as it is argued that it is not instructive to distinguish between sonifications and music/sound art. 

Edgard Varèse defined music as organised sound, and sonifications organise sound to reflect mimetically the thing being sonified. Therefore, an \textit{æsthetic perspective space} onto which sonifications and musical compositions alike can be mapped is proposed. The resultant map allows sonifications to be compared with works in the ars musica domain with which they share characteristics. The æsthetics of those ars musica counterparts can then be interrogated revealing useful design and organisation constructs that can be used to improve the sonifications' communicative ability.
\end{abstract} 
\section{Sonification and Æsthetics}
Whilst there are several terms used to describe the precise process by which data is rendered in sound, sonification is generally used as a catch-all to describe most work in the area of auditory display. Data can be mapped to sound in one of two ways: direct mappings impose a one-to-one relationship between data items and sonic events (possibly involving some scaling and quantisation) whilst metaphoric or analogic mappings impose interpretive filters or mapping functions to the data before it is rendered (Kramer \cite{Kramer:1994b} uses an analogic--symbolic continuum to classify audtory displays). An example of the former is Chris Hayward's auditory representation of seismograph data in which long period seismic waves were scaled up into the audible frequency range \cite{Hayward:1994} allowing sophisticated aural analysis. Blattner, Greenberg, and Kamegai \cite{Blattner:1992} used metaphoric mappings in their work on the auditory representation of turbulence in fluid flow. Through the use of earcons they represented changes in fluid state which would be hard to spot in a graphical display. Gaver's auditory icons \cite{Gaver:1989} are also interpretive but unlike earcons, auditory icons take a symbolic or representational (mimetic) approach. For example, a progress bar could be represented by the sound of a jug being filled: the fuller the jug sounds the nearer to completion the task is. 

At first, the focus in auditory display was to show how information could be mapped to sound. Many systems were built for a wide variety of information types (such as stock market data, seismographs, program runtime behaviour, chemical spectra, DNA sequences, chaotic attractor functions, etc. -- see Kramer \cite{Kramer:1994c}). Sonification designers concentrated more on building systems and less on those systems' æsthetic qualities. In the early proof-of-concept stage this was understandable, but now that the field is beginning to mature this issue needs to be addressed. Over the last decade or so many different sonification examples have been built. What is apparent from listening to them is the wide variation in the quality of the æsthetics and the acoustic ecology\footnote{By acoustic ecology here we mean the internal ecology of the various sounds within the sonification. That is, we treat the sonification both as a real-world soundscape in its own right the acoustic ecology of which is jumbled, and as part of the wider real-world soundscape in which it is situated. Again, its sonic components may sit uneasily within the acoustic ecology of the host soundscape.} of the auditory output. Some designers adopted simple quantised data-to-pitch mappings that allowed crude chromatic pitch mappings, whilst others pursued ad-hoc frequency mappings. Others tried deliberately to use organising principles of tonal music to structure their work, which resulted in more æsthetically-coherent sonifications. Another camp explored the electroacoustic music tradition to use in their sonifications the more extended concept of what sounds are suitable for musical use deriving from musique concrète\footnote{In musique concrète and electroacoustic music conventional pitched tones are only a subset of spectro-morphologies within a much broader world of spectra \cite{Smalley:1986}. The reliance on architectures based around harmonic progressions of pitches is removed}. What marked out these latter two camps was the use of musically-literate people on the project teams. These teams believed that people with formal musical and artistic training could significantly improve the æsthetic qualities of the sonifications which, in turn, would increase the communicative and expressive capability of the auditory displays. Kramer \cite{Kramer:1994b} commented on the similarity in structure between sonification and music creation: sonification renders data in sound to allow a human listener to detect and comprehend patterns and structures in that data, whilst a musician renders a musical score so as to make it audible and thus make perceptible the music's structure and even give clues as to the composer's and the musician's emotional states. We could go as far as to claim that a piano is a sophisticated auditory display machine: through the intervention of the musician, the piano renders in sound (albeit in a highly complex and abstract way) the score, the technique of the musician, the physics of the piano, the emotional state of the musician and the composer, and even the musician's response to the feedback loop offered by his own ears.

\section{Hearing and Listening}
In looking for easily understood mappings for his {\sc SonicFinder}, Gaver \cite{Gaver:1989} proposed a theory of `everyday listening' which says that in everyday situations people are more aware of the attributes of the \textit{source} of a sound than the attributes (parameters) of the sound itself: it is the size of the object making the sound, the type of the object, the material it is made of, etc. that interests the everyday listener. We hear big lorries, small children, plastic cups being dropped, glass bottles breaking, and so on. Everyday listening is in contrast to what Gaver calls `musical listening' in which we are more interested in attributes of the sounds themselves: their pitch, their intensity, and so on.  

Gaver's ideas are indebted to Schaeffer's \textit{Quatres Écoutes}, or `four modes of listening'.\footnote{The four being Causal Listening, Reduced Listening, Semantic Listening, and Technological Listening.} Gaver's `musical listening' resembles Schaeffer's \textit{écoute réduite} or Reduced Listening, though it is not entirely the same thing -- see Smalley \cite[p. 63]{Smalley:1986}. His `everyday listening' is very close to Chion's Causal Listening \cite{Chion:1994} (a term proposed as a contrast to Schaeffer's Reduced Listening). Of particular interest here is Chion's Semantic Listening in which the listener seeks to gain information about what the sound is communicating: we know it here as sonification!

However, it is not clear that Gaver's two types of listening are strictly categorical -- they would seem more to be points along a continuum describing general listening approaches but sharing attributes. For example, in musical listening, though one may be primarily interested in the harmonic and temporal relationships of sounds, the individual timbres and their sources are also an important part of the experience. There are `big' sounds and `little' sounds even in music.

Gaver's work was strongly motivated by R. Murray Schafer's concept of acoustic ecology in soundscapes (see \cite{Schafer:1977}). Schafer sees the world around us as containing ecologies of sounds. Each soundscape possesses its own ecology, and sounds from outside the soundscape are noticeable as not belonging to the ecology. In Schafer's worldview we are exhorted to treat the environments in which we find ourselves as musical compositions. By this we are transformed from being mere hearers of sound into active and analytic listeners -- exactly the characteristic needed to benefit most from an auditory display. When the environment produces noises that result from data and events in the environment (or some system of interest) we are able to monitor by listening rather than just viewing. This acoustic ecology viewpoint cuts across Gaver's everyday listening--musical listening divide as Schafer sees the everyday world as a musical composition. In so doing he brings together into a single experience Gaver's separate acts of everyday and musical listening -- we attend to the attributes of the sounds and the attributes of the sound sources equally\footnote{In fact, this is exactly what some electroacoustic composers strive to do – cut across the everyday and the musical, where the everyday sounds, with their indexical qualities, become musical, and bring into music the everyday.}; the way we listen to each sound is contextually dependent on the information we wish to draw from it. 
\subsection{Æsthetics and acoustic ecology}
Some have enthusiastically grasped notions of acoustic ecology in their sonification designs. In recent years there has been a growing realisation of the important role to be played by æsthetics in the design of computer systems and artefacts. Fishwick \cite{Fishwick:2002,Fishwick:2006} coined the term \textit{Æsthetic Computing} to refer to the application of art theory and practice to the design of computing systems. Fishwick claims that ``\textit{…there is a tendency toward the mass-media approach of standardized design, rather than an approach toward a more cultural, personal, and customized set of æsthetics}'' \footnote{Personal communication}. It is making use of cultural and personal differences that Fishwick claims will enlarge ``\textit{…the set of people who can use and understand computing.}'' Whilst Fishwick has focused primarily on the æsthetics of visualisations and models\footnote{The role of sonification in Æsthetic Computing has not gone unnoticed by Fishwick though as evidenced by the inclusion of a chapter by Vickers and Alty \cite{Vickers:2006} on program sonification in the volume Æsthetic Computing.}, the auditory display community has also begun to pay attention to æsthetic issues in sonification.

Cohen \cite{Cohen:1994} used acoustic ecologies of sounds in his {\sc ShareMon} system, only he called them collections of `genre' sounds. A principle of Æsthetic Computing is that systems should be malleable according to the culture in which it is situated \cite{Fishwick:2002}. Cohen argued the strongest reason for not using `genre' sounds is that they are less universal than everyday sounds (as used in auditory icons) or musical motifs (as used in earcons). He put the counter-argument thus: ``\textit{\ldots everyday sounds vary for different cultures anyway, as do the ways of constructing musical motifs}'', so there is no reason in principle why these types of acoustic ecology cannot be successfully used in the right context. Cohen also identified the importance of allowing users to assign their own choice of sound sets to auditory monitoring applications: this catering for user preference is another principle of Æsthetic Computing.  Indeed, Cohen sugggested that users could ``\textit{\ldots choose familiar genres based on æsthetic preference}'' \cite{Cohen:1993}.
\subsection{Musique Abstraite vs Musique Concrète: Tonal, Atonal, Serialist, Electro-Acoustic, \& Non-Tonal Music}

Drawing on the ideas of futurist composer Luigi Russolo (1885-1947), principles of Pierre Schaeffer's musique concrète, and inspired by Edgard Varèse's \textit{Poème Électronique} (1958) and John Cage's aleatoric compositions (e.g. \textit{Music of Changes} (1951)), Barra et al \cite{Barra:2002} tried to construct sonifications for monitoring a web server that were ``\textit{…neutral with respect to the usual and conventional musical themes.}'' They attempted to move away from the idioms of tonal and atonal (serialist) music (musique abstraite, or abstract music) and towards the more concrete compositions found in the musique concrète \& electroacoustic traditions. Musique concrète approaches composition not by writing a tune which is then given to players to render in sound\footnote{This traditional way of compusing music is known as musique abstraite. In this paper the term \textit{abstract} is used exclusively to refer to music of this type, rather than the more culturally prevalent usage which would describe electroacoustic music as `abstract' in the same sense that much modern art is abstract.} but instead by first recording existing or `found' (or rather, `chosen') sounds and assembling them into a musical piece. Vickers \cite{Vickers:2004a} sounded a note of caution at such moves away from tonal music systems:

\begin{quotation}In the pursuit of æsthetic excellence we must be careful not to tip the balance too far in favour of artistic form. Much current art music would not be appropriate for an auralisation system. The vernacular is popular music, the æsthetics of which are often far removed from the ideals of the music theorists and experimentalists \cite{Vickers:2004a}.
\end{quotation}
With hindsight this argument seems too simple as it embodies the thinking of C. P. Snow's \textit{Two Cultures} by creating a division between so-called art music and sonification. Vickers' position is based on the assumption that much art music would not be perceived as music by the average listener and, therefore, its structures would not be comprehensible. This view is supported by Lucas' observations \cite{Lucas:1994} that the recognition accuracy of an auditory display was increased when users were made aware of the display's musical design principles. Watkins and Dyson \cite{Watkins:1985} demonstrated that melodies following the rules of western tonal music are easier to learn, organise cognitively, and discriminate than control tone sequences of similar complexity. This, Vickers argued, meant that the cognitive organisational overhead associated with atonal (Schoenbergian serialism) systems makes them less well suited as carriers of information. However, this reasoning fails to account for the fact that electroacoustic music and musique concrète, whilst often lacking discernible melodies and harmonic structures, is still much easier to organise and decompose cognitively than atonal pieces \cite{Vickers:2005b}. The studies of Lucas \cite{Lucas:1994} and Watkins \& Dyson \cite{Watkins:1985} were rooted in the tonal/atonal dichotomy and did not consider these other branches of music practice. Vickers' revised opinion (see \cite{Vickers:2005b}) that the electroacoustic and musique concrète traditions can, in fact, offer much to sonification design supports the position taken by Barra et al in their design of the {\sc Webmelody} web server sonification system (see \cite{Barra:2002}). Indeed, Barra et al avoided the use of harmonic tonal sequences and rhythmic references because they believed that such structures might distract users by drawing upon their individual ``\textit{…mnemonic and musical (personal) capabilities}'' \cite{Barra:2002}. Rather, they ``\textit{\ldots let the sonification's timbre and duration represent the information and avoid recognizable musical patterns}''. This, they said:
\begin{quotation}
 \ldots makes it possible to hear the music for a long time without inducing mental and musical fatigue that could result from repeated musical patterns that require a finite listening time and not, as in our case, a potentially infinite number of repetitions \cite{Barra:2002}.
 \end{quotation} 
The electroacoustic and musique concrète approaches may potentially lead to great success in sonification design given their dependence upon the notion of `gesture' encoded into sounds. Smalley's classification of sounds as spectro-morphologies \cite{Smalley:1986} is based in part in the notion that we hear the physical, gestural qualities in sounds, and that these in and of themselves, though of course usually in combination with timbre and volume, carry enough information regarding movement, atmosphere, size, material quality, and so forth to offer information to the perceiver that serves to generate meaning akin to that generated by the musical harmonic/tonal system. It has the added advantage of being arguably less culture specific, that is it is not classical, or pop, or anything we already recognize -- it is rather a system that is more open to reading than it is a musical style that is recognized as such. 

For their {\sc Audio Aura} system (a monitoring system to allow people to have background awareness of an office environment), Mynatt et al \cite{Mynatt:1997a,Mynatt:1998} created four separate `ecologies' (their word) of auditory cues. The sounds within each set were designed to be compatible with the others not just in terms of frequency and intensity balance but in logical (and semantic) terms too. For example, their `sound effects world' ecology was based around the noises to be heard at the beach: gull cries were mapped to quantities of incoming email with surf and wave noises representing the activity level of members of a particular group. Thus, each sound in a particular ecology would not sound out of place with the others. In all, four ecologies were constructed:
\begin{enumerate}
\item Voice world -- vocal speech labels
\item Sound effects world -- beach noises: an auditory icon \& sound\-scape set
\item Music world -- tonal musical motifs: a structured earcon set
\item Rich world -- a composite set of musical motifs, sound effects, and vocal messages.
\end{enumerate}
Unfortunately, no formal studies have been published to suggest how well the ecologies worked and which of the four was better received by users. In theory, this selection of different ecologies allows user preference to be catered for which is an important principle in Æsthetic Computing \cite{Fishwick:2002,Fishwick:2006}. Such principles can also be found in Tran and Mynatt's {\sc Music Monitor} \cite{Tran:2000} which allowed the user to personalise the system by specifying their preferred music tracks upon which the main earcon messages were overlaid.
\subsection{Music Æsthetics}
One criticism levelled against using musical æsthetics in sonifications is that the musical grammars add another language level to the interface which would get in the way of the underlying data -- the music would be another language to learn. However, applying the argument to external visual representations shows it to be fallacious. Structured external visual representations are common in the computing world. The diagrams themselves are graphical abstractions of the underlying data or concepts they represent. No complaint is made that the syntax (or organising rules) of the notations interfere with understanding what they represent. Rather, it is considered necessary to have formalised rules by which diagrams and other notational structures are organised. It is held by some that musical/sonic syntaxes somehow require greater cognitive load than visual representations. There are two things being missed here. First, even visual notations require training in how to read them. Secondly, people are already familiar with decoding the organising principles of at least one sonic grammar. If it were not so, it would be impossible to appreciate music without formal training, yet melody recall has been shown to be an innate skill. Popular music would not exist unless it could communicate its message to a wide population with the minimum of cognitive overhead. It is true that people differ in the analytical level of their listening, but there does seem to be a cultural, or æsthetic, baseline in popular music systems that is accessible to the untrained listener.
\subsection{From ars musica to ars informatica and back again}
\begin{figure*}[hbt] 
\centering
\includegraphics[width=5 in]{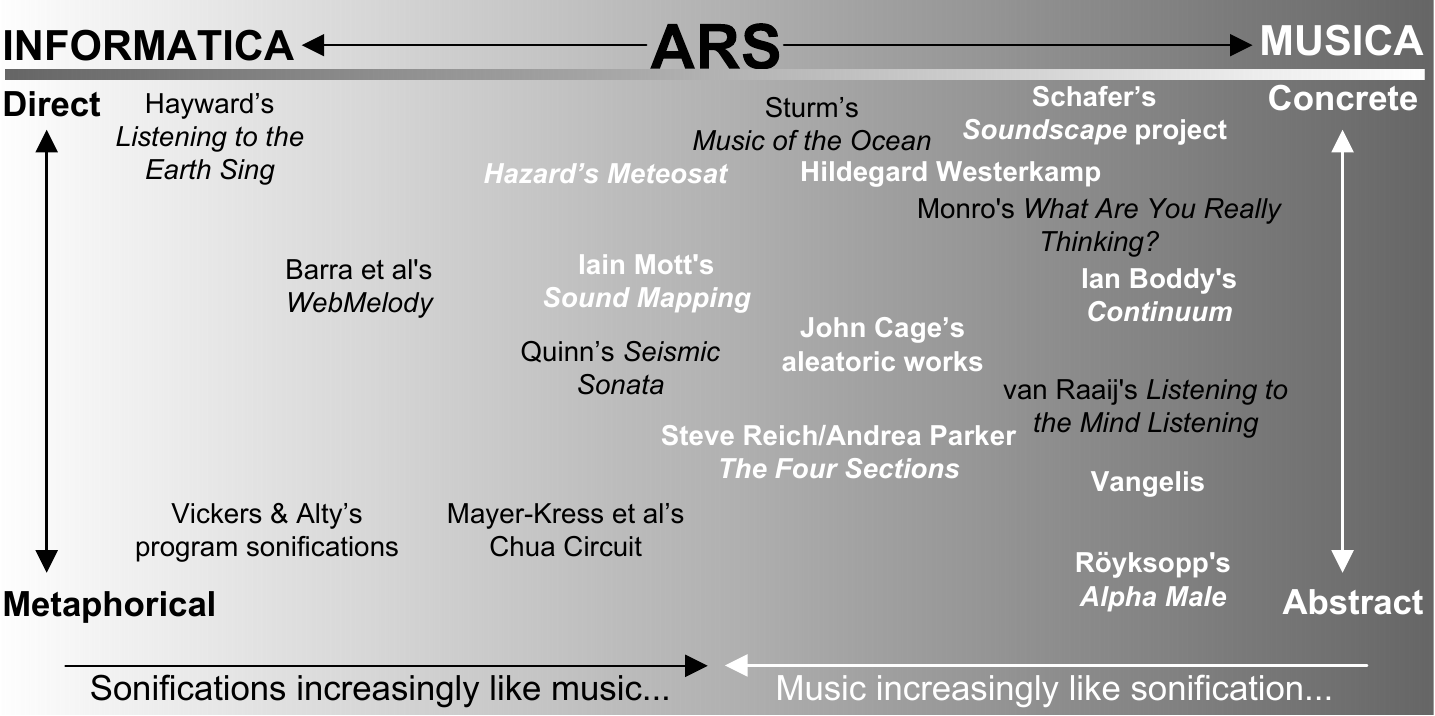} 
\caption[hello]{{\it An Ars Informatica--Ars Musica Continuum}}   
\label{fig:continuum} 
\end{figure*}

To move us towards how to think about the relationship between sonification and music æsthetics we may imagine a line, a continuum, with sonification (or, ars informatica) at one end and music and sound art (ars musica) at the other (Fig. \ref{fig:continuum}). At the ars informatica end lie sonifications without pretense to artistic content (though whether they are perceived as such is another matter): their intent is to create as pure a mapping from data to sound as possible\footnote{There is, in fact, a movement dedicated to so-called `pure data' mappings in which only the data are heard with no interpretive, abstraction, or quantisation layers imposed on top. However, we cannot see how this is achievable because the very act of transforming data to sound requires some form of mapping which is unlikely to be isomorphic.}.
At the ars musica end are those pieces of music and sound art that exist as pure art forms whose purpose is not necessarily to communicate discrete or quantitative data from the real world. Of course, it is debatable whether any such pieces exist, for all composers and creators of art typically try to communicate something in their work no matter how abstract\footnote{Behaviourists might take issue with this stance.}; however, this classification is a useful abstraction for our purposes here.

The continuum has a second dimension, the degree of \textit{indexicality}, that is, how strongly a sound sounds like the thing that made it. In the ars informatica end of the continuum, indexicality is related to whether sonifications make more use of direct data-to-sound mappings (high indexicality -- the sound is derived directly from the data) or more use of metaphoric or interpretive mappings (low indexicality). Hayward's auditory seismograms are an example of the former in which seismographic data were pre-processed with amplitude scaling, DC removal, and interpolation and then frequency-doubled, time-compressed, and and amplitude scaled (using automatic gain control) until they lay in the human audible range. Vickers \& Alty's program sonifications are metaphoric: tonal musical motifs were used to stand for data and objects. 

At the ars musica end of the continuum pieces are similarly categorised with the concrete works based on assemblies of found sounds and environmental sounds lying at the high indexicality end and the interpreted performances of `traditional' musical scores lying at the abstract (low indexicality) end. As direct sonifications and concrete music both possess high indexicality and metaphorical sonifications and abstract music low indexicality, we can suggest in more formal terms: 
\begin{equation} 
\begin{array}{l}
Direct \mapsto Concrete\\
Metaphorical \mapsto Abstract\\
\end{array}
\label{eq1} 
\end{equation} 
that is, `direct' and `metaphorical' in the sonification domain map to `concrete' and `abstract' respectively in the ars musica domain.

The interesting area lies in the middle where sonifications have been deliberately designed with artistic sensitivities (e.g. Mayer-Kress et al's \cite{Mayer-Kress:1994} sonification of a chua circuit chaotic attractor function) and where music has been composed according to some underlying data or process (e.g. John Cage's \textit{Music of Changes} (1951)). The line between sonification and music is blurred when we consider shared composition techniques. For example, Hayward's \cite{Hayward:1994} frequency doubling of seismographic data is analogous to the work of Hildegard Westerkamp\footnote{See \url{www.sfu.ca/~westerka/}} who \textit{slows down} environmental sounds to extract their previously unheard musical characteristics. They both manipulate the speed of the data but the intended outcomes are different.
\begin{figure*}[ht] 
\centering
\includegraphics[width=5.5in]{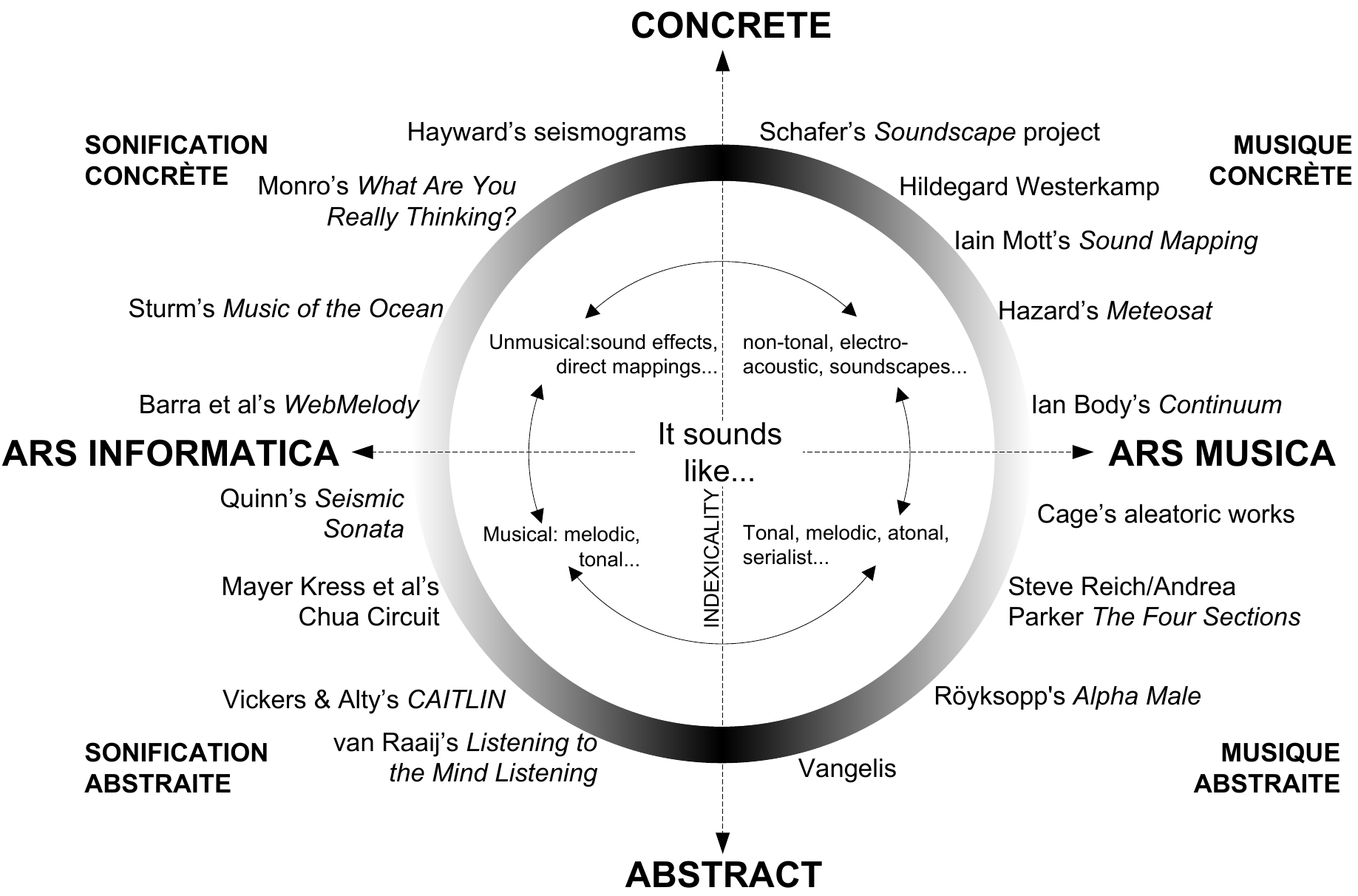} 
\caption{{\it The Ars Informatica--Ars Musica Æsthetic Perspective Space}}   
\label{fig:circle} 
\end{figure*} 

The melting pot is further stirred by sonifications that were designed as music. Consider Quinn's \textit{Seismic Sonata}\footnote{Marty Quinn, ``Seismic Sonata: a Musical Replay of the 1994 Northridge, California Earthquake'', 2000} or Sturm's \textit{Music from the Ocean}\footnote{Bob Sturm, ``Music From the Ocean'', composerscientistrecordings (\url{www.composerscientist.com}), 2002} which aimed to provide a sonification within a musical experience. Quinn maps multivariate earthquake data to sonata form; Sturm's sonifications of ocean buoy data are concrete, resembling Hazard's \textit{Meteosat}\footnote{Hazard, Fennesz, \& Biosphere, ``Light'', Touch Records, 2001} and Ian Boddy's \textit{Continuum}\footnote{Ian Boddy, ``Continuum'', Something Else Records, 1996, a two-CD abridged recording of a live eight-hour performance of electronic music by Ian Boddy at the Newcastle Comic Arts Festival in 1996}, yet with deliberate and planned mappings. More extreme examples can be heard in Monro's \textit{What Are You Really Thinking?} and van Raaij's \textit{Listening to the Mind Listening}. Both sonifications lie at the ars musica end of the contiunuum being designed as performance pieces for a concert (see discussion in section \ref{perspective} below) yet they are also sonifications of a 15-dimensional data set. Depending on one's perspective they can be listened to as sonifications or as pieces of music.

Rather than sonifications and compositions lying on a continuum we propose, instead, a circular (or spherical) space in which the ends of the continuum have been joined (Fig. \ref{fig:circle}). The right hemisphere represents the musical composition space, the left hemisphere the sonification space. The north-south axis (labelled concrete and abstract) denotes the indexicality of the pieces. In the south-east quadrant we place those compositions that sound most like tonal and atonal music (musique abstraite), and at the opposite point those sonifications that sound (to the layman) least like music or possess the least musical quality (sonification concrète), that is, those sonifications that use the most direct data-to-sound mappings. This division between concrete music and least-musical sonifications is contentious for it requires a consensus as to what constitutes music in the first place, but for now it is a useful distinction that we will shortly discard or rather, transform. In the north-east quadrant lie the non-abstract compositions (musique concrète) with the south-west being occupied by the more melodic sonifications (sonification abstraite, or those sonifications that make use of indirect and metaphorical data mappings).  

\subsection{Perspective}\label{perspective}
As we travel around the circle in either direction we meet those works that possess to a greater or lesser extent attributes of both musical composition and non-musical sonification. The really interesting point is where it is hard (or even impossible) to discern the origins of a piece: composition or sonification. This implies that the distinction between the two pairs of indexicality polarities -- musical (tonal) sonification \& abstract music (bottom hemisphere) and non-musically-sounding sonification \& concrete music (top hemisphere) -- become blurred. By shifting perspective, we can transform the way we listen so that an unmusical sonification becomes a piece of musique concrète, and a musical sonification becomes a piece of musique abstraite (and vice versa). Indeed, musicologists would argue that music is as much a construct of the listener’s mind as a construct of the composer; if the listener perceives something as music then it is music(though Smalley \cite{Smalley:1986} claims that the listener must  ``\textit{…discover a perceptual affinity with its materials and structure}'' in order for this to happen). If that is so we can state:
\begin{equation} 
Sonification \implies Music
\label{eq2} 
\end{equation}
that is, if we have a sonification then we also have music. By the same token, we may consider any piece of music as a sonification giving:
\begin{equation} 
Music \implies Sonification
\label{eq3} 
\end{equation}
that is, if we have a piece of music then we also have a sonification. These two statements, therefore allow us to note:
\begin{equation} 
Sonification \iff Music
\label{eq4} 
\end{equation}
and say that whether we hear a sonification or a piece of music is simply a matter of perspective. By way of example, consider the \textit{Listening to the Mind Listening} concert at ICAD 2004.\footnote{Binaural recordings of all ten works, and the EEG data on which they are based, are available to download at \url{www.icad.org/websiteV2.0/Conferences/ICAD2004/concert.htm}} The concert comprised ten five-minute sonifications of fifteen channels of EEG data previously recorded from a subject who was listening to David Page's \textit{Dry Mud}\footnote{David Page, ``Dry Mud'' from the score of ``Fish'', (moving picture, Australian National Film and Sound Archive no. 413270), 1999 }. The compositions were driven by the same data but the mappings were defined by their creators. Gordon Monro's \textit{What Are You Really Thinking}\footnote{\url{www.icad.org/websiteV2.0/Conferences/ICAD2004/concert/audio/Monro-binaural.mp3}} sounds very much like an electroacoustic composition. At the other end of the spectrum (on the other side of the circle) is Hans van Raaij's \textit{Listening to the Mind Listening}\footnote{\url{www.icad.org/websiteV2.0/Conferences/ICAD2004/concert/audio/VanRaaij.mp3}} in which the data were mapped onto just two auditory streams, a piano and bass, creating a kind of tonal free jazz improvisation, a very different æsthetic form from Monro's work. By considering the two pieces in terms of indexicality and musicality we see how our perspective has changed thus moving the works from the sonification side of the space into the composition side. The difference, then, between sonification and musical composition is largely one of perspective. The ars informatica--ars musica continuum (Fig. \ref{fig:continuum}) was useful as a starting point but can now be left behind in favour of the Æsthetic Perspective Space (Fig. \ref{fig:circle}) in which the main dimension is indexicality (or the musique abstraite--musique concrète spectrum). If we imagine the perspective space as a sphere with `concrete' and `abstract' lying at either pole of a rotational axis, then just as east meets west on the earth, so travelling far enough towards `ars informatica' brings one eventually into the `ars musica' perspective domain. Another way of considering the perspective space is to fold it vertically along the concrete--abstract axis so that all sonifications are now thought of as compositions.

We acknowledge that the semantics of Figs. \ref{fig:continuum} and \ref{fig:circle} are not greatly different from each other as they share the same dimensions. However, Fig. \ref{fig:circle} is an attempt to remove the implied distance between the two ends of the continuum: in Fig. \ref{fig:continuum} it appears that sonifications and compositions at either end of the continuum are more different from each other than sonifications and compositions towards the middle. Fig. \ref{fig:continuum} highlights the categorisation along the ars informatica--ars musica dimension. However, we have argued that any difference along this dimension can be viewed largely as one of {\it perspective}, that any piece of ars informatica can also be thought of as ars musica and vice versa. Fig. \ref{fig:circle}, then, attempts to remove this dichotomy by placing the pieces on a circle so that the ends of the continuum are not semantically distant. The vertical placement (indexicality) of the works is also a rough attempt at indicating musical similarity between the left and right hemispheres (q.v. section \ref{interrogation}): works of similar latitude share musical characteristics and features, a phenomenon which is clearly seen when the space is folded along the indexicality axis which causes works at the same latitude to overlay each other. Longitudinal placement in Fig. \ref{fig:circle} is not intentionally significant. Works have been set on the perimeter of the circle for ease of diagrammatic layout. A three-dimensional spherical arrangement would allow greater flexibility of exploring the significance of true longitudinal placement.

\subsection{Æsthetic Interrogation}\label{interrogation}
By folding the perspective space along the abstract/concrete axis we place the sonifications into their corresponding genre (indexicality) level in the ars musica domain. This enables an æsthetic interrogation of the different compositional techniques to elicit the fundamental organising principles and æsthetic properties of various ars musica genres. Indeed, this transformation shows the structural similarities of certain pairs of ars informatica---ars musica quite clearly, e.g. Sturm's \textit{Music from the Ocean} \& Hazard's \textit{Meteosat} and Hayward's seismograms \& Schafer's soundscapes. These principles can then be applied to sonifications and thus, it is hoped, strengthen their æsthetic and communicative properties.

\section{Conclusions and Reflections}
To distinguish between musical and non-musical sonifications is not necessarily helpful (or even meaningful). Wrightson \cite{Wrightson:2000} invokes R. Murray Schafer \cite{Schafer:1977} who ``\textit{\ldots suggests that we try to hear the acoustic environment as a musical composition and further, that we own responsibility for its composition}''. A reason for trying to hear everything as a musical composition is that it forces us to move from being hearers to being listeners. What then becomes important for sonification designers is not how `musical' their work sounds, but how easy they have made it for the audience to listen to it, and by listen we mean `attend carefully'. Timbres should be chosen so that they do not mask other timbres (unless masking is an important feature of the underlying data). Careful thought needs to be given to the spatialisation of the sounds -- mono, stereo, and multi-channel sound can all be used to good effect but only if used well.

Researchers who have reported the most success also tended to deal directly with the issue of the æsthetics and acoustic ecology of their sonifications. As the role of æsthetics is increasingly entering the consciousness of designers of computing systems, so it needs to inform the work of the auditory display community. It is proposed that sonifications be viewed as works of ars musica as they could then benefit from the application of the æsthetic practices employed by artists. The sonifications discussed in this paper have been primarily for task monitoring (Hayward's seismograms excepted). In this mode of listening the æsthetics play an important role in reducing fatigue and annoyance. Just as poor æsthetics can get in the way of listening, `good' æsthetics might become too seductively musical, distracting the listener from the information that is to be conveyed (recall St Augustine's confessions, seeking forgiveness from God for having found more pleasure in the music than the texts sung). This danger is potentially greater in data set analysis tasks (such as data mining or seismographic analysis) in which the fine grain detail is much more important than in process monitoring tasks.

The foremost skill that a sonification designer needs to develop is that of listening, for it is upon this that all higher sound art skills are predicated. Once sonification designers have learnt to listen like composers, sound designers, and recording engineers they will be much better placed to create sonifications that maximise the communicative potential of the auditory channel. Designers of visual interfaces have been drawing upon the skills of the graphic design community for years: it is time the auditory display community did something similar.

The argument in this paper is deliberately centred around those attributes that music and sonification share, as it is at these intersections that dialogue and interrogation may take place. There are other artefacts present in each of music and sonification that are not present in the other, and we have not tried here to assess how these other attributes affect æsthetic interrogation. One such aspect is the intellectual content of compositions. Musical compositions are deliberate works of intellect and when the primary motivation is musical expression and experience (rather than communication of data), there is an intellectual component to the music that sits beyond the scope of the discussion in this paper. One way of approaching the impact of intellectual content of sonification--music may be to  use structures such as Emmerson's {\it Language Grid} \cite{Emmerson:1986} in which the syntactic abstraction of music may be categorised and analysed. 

The `Æsthetic Perspective Space' is a useful starting point for commencing the development of a framework for æsthetic interrogation in order to apply æsthetic principles of ars musica to ars informatica and it facilitates the application of electroacoustic spectro-morphological thinking to the design of sonifications. 

\bibliographystyle{IEEEbib} 
\newcommand{\noopsort}[1]{} \newcommand{\singleletter}[1]{#1}

\end{document}